\documentclass[]{spie}  %>>> use for US letter paper
\usepackage{amsmath,amsfonts,amssymb}
\usepackage{graphicx}
\usepackage[colorlinks=true, allcolors=blue]{hyperref}

\title{Planet Formation Imager (PFI): Project update and future directions}

\author[a]{John D. Monnier}
\author[b]{Stefan Kraus}
\author[c]{Michael J. Ireland}
\affil[a]{University of Michigan, Ann Arbor, MI USA}
\affil[b]{University of Exeter, Exeter UK}
\affil[c]{Australia National University, Canberra Australia}

\authorinfo{Further author information: (Send correspondence to J.D.M.)\\J.D.M.: E-mail: monnier@umich.edu}

\begin{document} 
\maketitle

\begin{abstract}
The Planet Formation Imager (PFI) Project is dedicated to defining a next-generation facility that can answer fundamental questions about how planets form, including detection of young giant exoplanets and their circumplanetary disks.  The proposed expansive design for a 12-element array of 8m class telescopes with $>$1.2 km baselines would indeed revolutionize our understanding of planet formation and is technically achievable, albeit at a high cost. It has been 10 years since this conceptual design process began and we give an overview of the status of the PFI project.  We also review how a scaled back PFI with fewer large telescopes could answer a range of compelling science questions, including in planet formation and as well as totally different astrophysics areas.
New opportunities make a space-based PFI more feasible now and we give a brief overview of new efforts that  could also pave the way for the Large Interferometer for Exoplanets (LIFE) space mission.
\end{abstract}


\section{INTRODUCTION}
The Planet Formation Imager (PFI) project was initiated in 2013 following an interferometry workshop at the Observatoire de Haute Provence.  Over the years, the PFI Project (directed by John Monnier) has grown to include a Science Working Group (SWG, led by Stefan Kraus) and a Technical Working Group (TWG, led by Michael Ireland) involving around 100 scientists and engineers from around the world.  Various scientific and technology aspects of PFI have been explored already in a series of papers first at the 2014 SPIE
\cite{pfimonnier2014,pfikraus2014,pfiireland2014}, the 2016 SPIE
\cite{pfimonnier2016,pfikraus2016,pfiireland2016,pfiminardi2016,pfimozurkewich2016real,pfibesser2016,pfipetrov2016}, and 2018 SPIE \cite{pfi2018,2018pfi1,2018pfi2,2018pfi3,2018pfi4}.  Some of the high-level conclusions were subsequently published in a refereed article \cite{monnier2018_pfi} and an Astro2020 Decadal Survey White Paper \cite{monnier2019_astro2020}.

The PFI Project Team concluded in 2018 that the primary PFI goal of mid-IR imaging of complex dust structures around young stars would require 12x8m telescopes while a more modest goal of detecting accreting giant exoplanets in disks could be done with 4x8m or 12x3m telescopes at L band.  While a technology development plan was identified, there has been limited prospect for future funding for a PFI facility so far.

While there has been little progress toward construction of  the PFI facility since 2018,  the science case for PFI has only strengthened since then.  We have seen the success of the VLTI 4x8m exoGravity\cite{lacour2021} project in detecting many exoplanets, the discovery of the accreting circumplanetary disks\cite{wang2020,benisty2021}  in PDS 70 (a new poster child for PFI ) and also the successful launch of JWST \cite{christiaens2024} which has shown the power of the mid-infrared.  

Here, we will re-summarize the basic goals and design of the Planet Formation Imager project as it was laid out in 2018.  Then we will review the recent strategy meeting sponsored by the UK Royal Astronomical Society in 2024 March  that was organized by Stefan Kraus.  Lastly, we will discuss some alternate paths for PFI, including formation-flying space interferometry and other potential facilities with different science goals.

More information on the PFI Project, along with reference documents and powerpoint slides, can be found at the project website \url{http://planetformationimager.org}, although it has not been updated significantly for a few years.

\begin{figure} [ht]
   \begin{center}
   \begin{tabular}{c} %% tabular useful for creating an array of images 
   \includegraphics[height=4in]{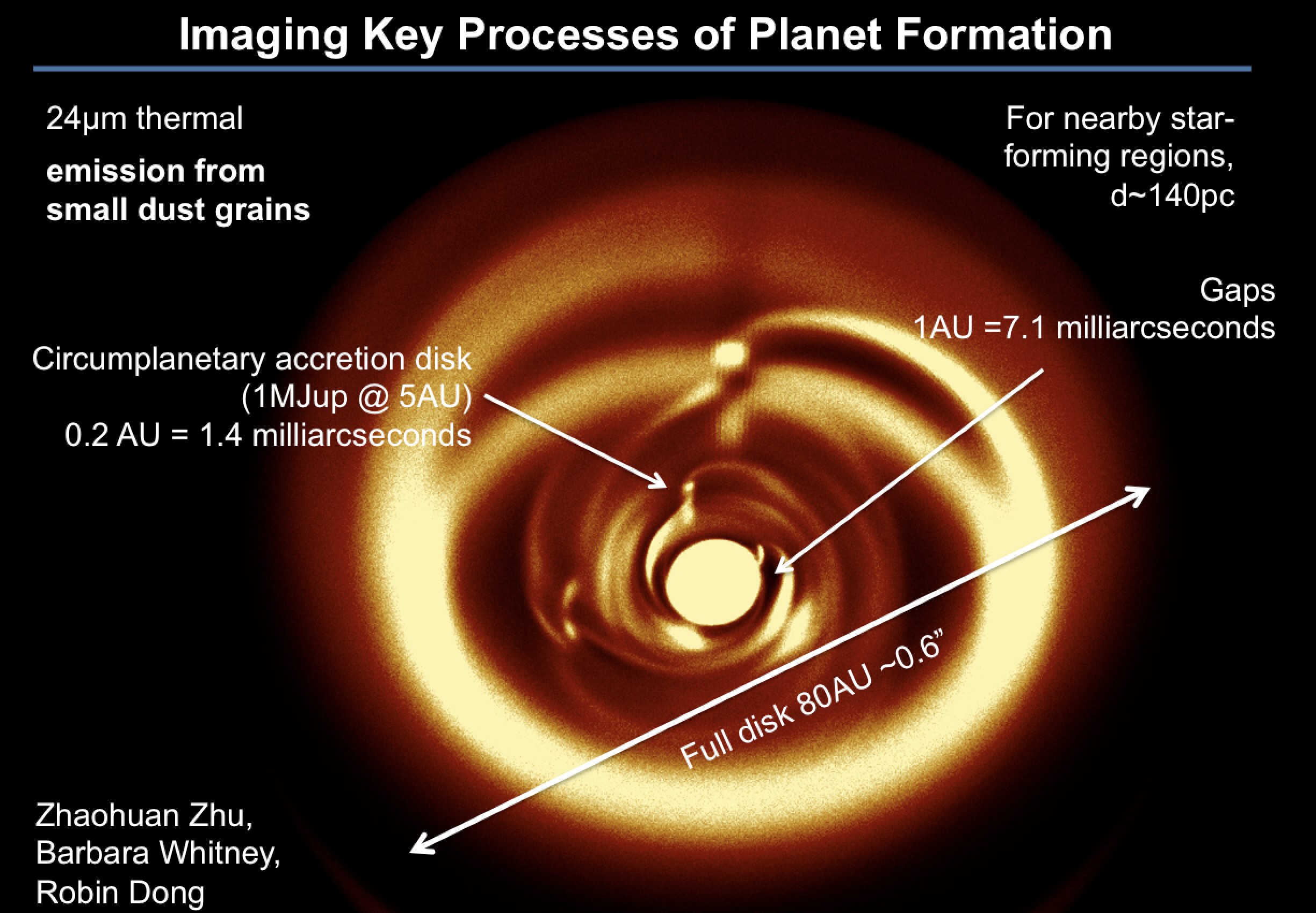}
   \end{tabular}
   \end{center}
   \caption[example] 
%>>>> use \label inside caption to get Fig. number with \ref{}
   { \label{fig:science} 
This figure shows the main planet formation processes that the Planet Formation Imager (PFI) Project was formed to address.  PFI should probe thermal dust emission to reveal gaps cleared by forming protoplanets (e.g., 1\,au gaps at 5\,au; 0.2\,au gaps at 1\,au) with enough resolution to just resolve circumplanetary disks themselves (0.2\,au for 1\,M$_{J}$ at 5\,au).  We also want to detect all the young and warm Jovian planets themselves throughout the roughly 80\,au disk.  
   }
\end{figure}

\section{Planet Formation Imager: Science Goals}

The foundation of PFI was to define a facility to image key stages of planet formation {\em in situ} down to the scales of individual circum-{\em planetary} disks and with sensitivity to characterize young giant exoplanets themselves. Detailed science cases were developed to define a series of ``top-level science requirements'' (TLSRs) before arriving at a facility architecture (see Table~\ref{tab:tlsr}).

\begin{table}[ht]
\caption{Typical absolute magnitudes for the emission components in protoplanetary disks \cite{pfimonnier2016,baraffe1998,spiegel2012,zhu2015,dong2015}, for the wavelength bands relevant for PFI, including adaptive optics system (Y band), fringe tracking system (H band), young exoplanets and dusty structures in the disk (L and N band).
To convert these absolute magnitudes to apparent magnitudes for an object located in Taurus at 140\,pc, simply add 5.7 magnitudes to the numbers below. }
\label{relevantfluxes}
\begin{center}       
\begin{tabular}{|l|c|c|c|c|}
\hline
Component & $M_Y$ & $M_H$ & $M_L$ &$M_N$ \\
          & (AO & (fringe & (dust & (dust \\
          & system)  & tracking) & \& planets) &  \& planets)\\
\hline
Example T Tauri Star & & & &  \\
\qquad 1 M$_{\odot}$, 2.1\,R$_{\odot}$, 3865\,K& 4.9 & 2.54 & $\sim$2.5  & $\sim$2.4 \\
\qquad 3\,Myr, [Fe/H]=0.0 &&&&\\
\hline
Protoplanet  & & & &  \\
%\qquad ``hot start" 10M$_J$ & & 8.7 & 7.8 & 7.5  \\
%\qquad ``cold start" 10M$_J$ & & 16.8 & 14.1 & 11.9   \\
\qquad ``hot start", 2M$_J$, 1\,Myr & & 12.9 & 11.0 & 9.1  \\
\qquad ``cold start", 2M$_J$, 1\,Myr& & 18.2 & 14.7 & 11.2   \\
\hline
Circumplanetary Disk & & & & \\
\qquad ($R_{in}=1.5~M_J$) & & & & \\
%\qquad  $ M\dot{M}=10^{-5}M^2_J\,{\rm yr}^{-1}$ & & 10.6 & 6.9 & 4.6  \\
\qquad $ M\dot{M}=10^{-6}M^2_J\,{\rm yr}^{-1}$ & & 16.4 & 9.8 & 6.5  \\
\hline
4-planet gapped disk & & & &\\
\qquad Star only (2\,R$_{\odot}$, 4500\,K)  & 4.1 & 2.1 & 2.1  & 2.1  \\
\qquad Star + Disk (30$^\circ$ inclination) & 4.1 & 2.1 &  1.6 & -1.1  \\
\hline
\end{tabular}
\end{center}
\end{table}

Figure~\ref{fig:science} gives a visual overview of the PFI science goals and relevant spatial scales.  We see PFI should have a field-of-view of at least 0.6'' to include the main portion of the planet-forming disk for nearby star forming regions. The angular resolution should be sufficient to not only resolve gaps caused by giant planet formation (e.g., 1\,au  gap at 5\,au) but also to resolve individual {\em circumplanetary} disks (e.g., 0.2\,au for 1\,M$_J$ at 5\,au). 
We have set a goal for PFI of imaging the dust around the 150\,K water iceline, as we expect this radius to be related to the zone of giant planet formation and includes the region of the disk where H$_2$O-rich asteroids form that eventually deliver water to terrestrial worlds\cite{raymond2017,walsh2011}.

The mechanism by which a young planet accretes dust and gas through circumplanetary disks is a poorly understood process and a key ingredient to planet formation theory \cite{ayliffe2009}. The size scale of the disk is expected to be about $\frac{1}{3}$ size of the Hill Sphere $R_H=a\sqrt[3]{\frac{m_p}{3M_\ast}}$.  Molecules\cite{rigliaco2015} could be present in the region around an accreting planet as well as in the disk itself, such as HI (7-6), H$_2$O, CO, CO$_2$, CH$_4$, C$_2$H$_2$, NH$_3$. Also exoplanets will leave an imprint in the   distribution of molecules in the disk, allowing for their discovery (e.g., as has been done recently with ALMA for very wide planets\cite{teague2018,pinte2018}).
Much more work\cite{ruge2014} is needed to understand how accreting protoplanets might be observable and with what tracers, and recent observations of PDS 70 are leading to new insights.

The last key science topic brought into the core PFI goals was to  detect directly and characterize all the giant planets younger than 10\,Myr around young stars.  Giant planets have relatively high temperatures after they form\cite{spiegel2012} and are even brighter while accreting\cite{zhu2015}, making them ideal targets for high angular resolution searches.  While ELTs will be able to detect some far-out giant planets (for reference, $\lambda/D$ at 3.5\,$\mu$m for ELT is 2.5\,au at Taurus, comparable to ALMA), an interferometer with kilometric baselines will be needed to  have  sufficient angular resolution to see planets within a few au and also will better resolve out dust features that can make it hard to spot young exoplanets  at these earliest stages. Because giant planets migrate or interact dynamically with their disk and also other planets, we want to measure any   significant differences in the location of giant planets at age 10\,Myr versus 100\,Myr and PFI should be sensitive to stars at the young ages when the gas disk is still important to processes such as migration, though young enough to be before most dynamical instabilities have been triggered.  Indeed, understanding giant planet formation is key to understanding terrestrial planet formation\cite{raymond2006b,baruteau2014,davies2014}.

Table~\ref{relevantfluxes} contains  some relevant fluxes of stars, young giant planets, and accreting protoplanets\cite{pfi2018}.  This information and the above science goals led to a set of ``top-level science requirements (TLSRs)'' and these are collected in Table~\ref{tab:tlsr}.

The philosophy of the PFI project had been to develop the science goals first then see what facility can meet those goals. After series of studies outlined in earlier SPIE papers,  our team has converged on a direct detection 1.5-13\,$\mu$m (H,K,L,M,N bands) long-baseline interferometer on the ground to best achieve the goals, although heterodyne detection and space interferometry were recognized as potential alternatives depending on technology progression.

\begin{table}
%\centering
\begin{center}
\caption{Top-level Science Requirements\label{tab:tlsr}
}
\begin{tabular}{|l|c|c|}
%\hline
%& Dust Imaging & Young Exoplanets \\
\hline
 Parameter & Dust Imaging & Young Exoplanets\\
\hline
Wavelengths & 5-13\,$\mu$m & 3-5\,$\mu$m \\
Typical Source Distance & 140\,pc & 50-500\,pc \\
Spatial Resolution & 2\,mas $\equiv$ 0.3\,au & 0.7\,mas $\equiv$ 0.1\,au (for 140pc)\\
%\qquad (extended goal)    & (0.7\,mas $\equiv $ 0.1~AU) & (0.2\,mas $\equiv$ 0.03~AU)\\ 
Point-Source Sensitivity & $m_N\sim12.5$ (5-$\sigma$)  & $m_L\sim18.5$ (5-$\sigma$) \\
\qquad (t$=10^4$s) & & \\
Goal Surface Brightness (K) & 150\,K  &  $--$\\
\qquad (t$=10^4$s) & & \\
Spectral Resolving Power&&\\
\qquad Continuum      & R$>100$  & R$>100$ \\
\qquad Spectral Lines & R$>10^5$ & R$>10^5$ \\
Field-of-view &  $>0.15"$  & $>0.15"$ \\
Minimum Fringe Tracking Limit & $m_H<9$ (star only) & $m_H<9$ (star only)\\
Fringe tracking star & $\phi<0.15\,$mas & $\phi<0.15\,$mas \\
\hline
\end{tabular}
\label{tlsr}
\end{center}
\end{table}

\section{Planet Formation Imager: Proposed Architecture}

Also in 2018, the Technical Working Group finalized reference facility architectures for use in designing detailed science cases.  We adopted the facility characteristics for our reference architectures:
\begin{itemize}
\item 1.2\,km maximum baseline chosen to attain 0.2\,au mid-IR resolution at 140\,pc: a) to resolve planet-opened gap for Jupiter 1\,au, b) to resolve diameter of circumplanetary disk for exo-Jupiter (1\,M$_{J}$@5\,au)
\item $12\times3$\,m diameter array chosen to have $T=150$\,K 3$\sigma$ surface brightness in 10000sec: 
Sensitivity to dust at $T=150$\,K, the temperature for the water iceline for typical disks
\item $12\times8$\,m diameter array chosen for enhanced surface brightness limit ($T=125$\,K) to see gaps and dust structures for giants planets forming as far out as 5\,au, and an enhanced exoplanet yield.
\item Sufficient fringe tracking margin (H band magnitude limit at least 13) to observe solar-type stars in nearby star forming regions, even with some extinction and visibility reduction by an inner disk.
\item Sufficient point source sensitivity to detect young giant exoplanets for a range of models.
\item Nuller design for exoplanet detections at K, L bands.  
\end{itemize}

Table\,\ref{pfiref} contains information on two reference facility architectures for PFI, one is a $12\times$3\,m array and the other a $12\times8$\,m array.  The main difference is a factor of 7 in sensitivity that is crucial for imaging warm dust at 5\,au and for a complete census of giant planets in young disks.  The notional construction costs contained in this table are based on estimates in [\citenum{pfiireland2016}] and with 2018 pricing\cite{kingsley2018} for $12\times$6.5\,m telescopes.

\begin{table}
%\centering
\begin{center}
\caption{Technical Description of Reference PFI Architectures\label{pfiref}}
\begin{tabular}{|l|c|c|}
\hline
Parameter & 12$\times$3m PFI & 12$\times$8m PFI \\
\hline 
%Priority Science & exoplanet-focused & dust \& exoplanets \\
%Optimised for cost.
Number of Telescopes & 12  & 12 \\
Telescope Diameter & 3\,m  & 8\,m \\
Maximum Baseline & 1.2\,km & 1.2\,km \\
Goal Science Wavelengths & 3--13\,$\mu$m  &3--13\,$\mu$m \\
Fringe-tracking wavelengths & 1.5--2.4\,$\mu$m & 1.5--2.4\,$\mu$m \\
Fringe tracking limits (point source) & $m_H<$13  & $m_H<$15 (AO-dependent)  \\
Point source Sensitivity (10$^4$s) & 18.1 (L), 12.2 (N) & 20.2 (L), 14.3 (N)   \\
Surface Brightness Limit (10$^4$s,$B=1.2$km) & 150\,K (N)& 125K (N) \\
Field-of-view & 0.25" (L), 0.7" (N) & 0.09" (L), 0.25" (N)\\
Note & w/ Nulling (2-4\,$\mu$m) &  w/ Nulling (2-4\,$\mu$m) \\
Construction cost & \$250M & \$600M$^*$ \\
\hline
\end{tabular}
\label{pfispecs}
\end{center}
{* Telescope cost based on informal estimate for 12$\times$6.5m telescopes\cite{kingsley2018}.}
\end{table}

\begin{figure} [hb!]
   \begin{center}
   \begin{tabular}{c} %% tabular useful for creating an array of images 
   \includegraphics[width=6in]{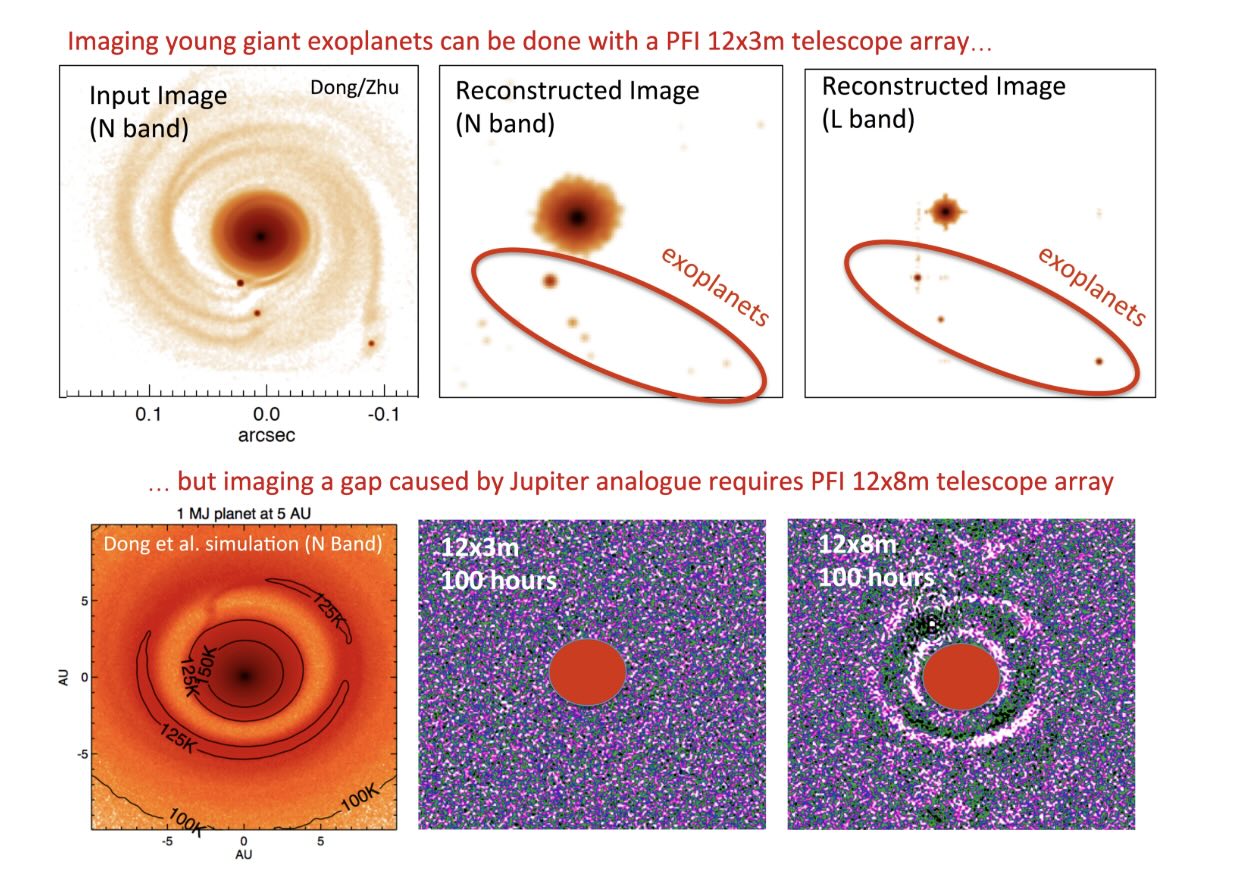}
   \end{tabular}
   \end{center}
\caption[example] 
{ \label{fig:newsim} 
The PFI Science and Technical Working Groups have simulated performance for both 12x3 m and 12x8 m
PFI architectures. (top row) Ref. \citenum{pfimonnier2016}  demonstrated that an equivalent 12x3 m PFI (in 25 hours) can detect
young giant exoplanets at both L and N bands (“hot start” models shown here) but cannot see the warm dust in this
4-planet simulation\cite{dong2015}. (bottom row) Ref. \citenum{pfi2018} found that a 12x8 m PFI array would
be able to image gap formation in solar analogues found in nearby star forming regions, a stunning capability bringing
ALMA-like performance to the inner solar system where terrestrial planets form.  Figure originally shown in Ref. \citenum{monnier2019_astro2020}}
\end{figure}

\section{RAS Meeting Summary: 2024 March}
\label{sec:ras}
In 2024 March, a community meeting was held as part of a UK Royal Astronomical Meeting to discuss a roadmap for a future kilometric baseline interferometric facility.  Updates on PFI were given by Stefan Kraus and Michael Ireland.  In addition, the group heard updates from other facilities:
\begin{itemize}
\item  CHARA's plan to extend to kilometer baselines using a mobile telescope, 
\item the possibility of a new 8m telescope for VLTI to give a $\sim$1 km baselines
\item the upcoming first fringes from the MROI
\item a new array of small telescopes being designed at Lowell Observatory
\item a moon-based interferometer (MOON-LITE)
\item progress on Intensity and Heterodyne Interferometry efforts
\end{itemize}

A more complete and comprehensive summary of the RAS meeting will be given elsewhere.

Michael Ireland led a discussion of potential high impact science questions that might be addressed by a kilometer-baseline interferometer in the optical and infrared.  Here are list of questions that came from this exercise:
\begin{itemize}
\item What is dark matter, dark energy, observational cosmology, distance scale, and inflation?
\item  What are the first stars?
\item  What are seed SMBH process/mass?
\item  How AGN and galaxies co-evolve?
\item  What are components of the baryon cycle and metal production
\item  What are GW, black hole and neutron star progenitors
\item  How does star formation quench, and what determines galaxy morphology
\item  Transients: what are they? (extreme physics)
\item  Galactic Archaeology of the MW. [X,H] dependent stellar evolution \& enrichment
\item How do planets form \& evolve?
\item Which planets are suitable for life?
\item Where is life amongst the nearest star systems?
\item How does the stellar dynamo evolve and what are the impacts on habitability?  How do starspots manifest themselves across the H-R diagram?
\item How do stars evolve for different metallicities, magnetic fields, spot coverages, and rotations rates?  How is angular momentum distributed in evolving rotating stars?
\end{itemize}

Interestingly, most of these science address questions by observing relatively simple, faint and small systems.  An array of 3 or 4 large ($>$8m) class telescopes with $\sim$1~km baselines operating in the visible or near-infrared would be best-suited.  This is quite a bit different than the original PFI design of 12x8m operating out to 10$\mu$m.

\section{New Paths to and from PFI}

\subsection{Space Interferometry}

Given breakthroughs in space launch vehicles (e.g., SpaceX) and increasingly sophisticated formation flying space missions, many in the PFI community have begun to reconsider space interferometry as a viable alternative to the 12x8m ground-based facility originally considered.  While a large imaging interferometer seems still far in the future, a successful smallSat interferometer would only require modest improvements to what have already been demonstrated\cite{monnier2019_roadmap,ireland2020}. Recently, M. Ireland (ANU) has started building the  Pyxis testbed (\url{www.mso.anu.edu.au/pyxis/}) that uses portable and moving robots to simulate 6-dof motions of CubeSat telescopes reflecting starlight to an immobile central station for true star tracking and interferometric beam combination.  Similarly, J. Monnier's group at Michigan is using drones to test long-baseline formation flying concepts (see contribution in these proceedings).  There has also been subsystem level description of low-power metrology systems \cite{lagadec2020} and novel nulling combiners suitable for space \cite{martinod2021}. Lastly, we note other independent ideas for feasible smallSat interferometers \cite{hansen2020,dandumont2021,matsuo2022} in low-earth orbit (and STARI project described by Monnier in these proceedings).

The Large Interferometer for Exoplanets (LIFE) mission\cite{quanz2022} is a re-boot of the Terrestrial Planet Finder Interferometer and DARWIN concepts from the 2000s. LIFE would be a mid-infrared nulling space interferometer that would be detect terrestrial planets around nearby stars, competitive with NASA's planned Habitable Worlds Observatory but at a lower cost.   The LIFE missions would be able to address many of the PFI science goals naturally, as LIFE's primary goal of detecting rocky Earths is already a more challenging goal than PFI's goal of finding young giant planets. 

\subsection{Forming Planets Interferometer (FPI)?}
Earlier studies by the PFI team and documented in refereed papers\cite{wallace2019}, found that the 4x8m VLTI could be a mini-PFI capable of detecting forming giant exoplanets in the K-L band range.  An additional 8m telescope at longer baselines would be helpful since the max baseline for VLTI-UTs is only 130m -- see SPIE paper by Bourdarot in these proceedings.   

Detecting exoplanets within the primary beam of an UT telescopes generally requires nulling and the VLTI-NOTT project\cite{Defrere:2018} will be the first serious attempt at nulling short of 10$\mu$m. Calculations show that many young planets will be detectable using the 4x8m VLTI-UT telescopes. See overview talk by Defr\`{e}re at this meeting for more information on VLTI-NOTT.

Perhaps smaller-diameter telescope arrays like CHARA or MROI could even detect exoplanets in H/K band using nulling, but this would require improvements to the infrastructure, including state-of-the-art adaptive optics systems and special combiners.  See talk by Monnier in these proceedings describing exoplanet searches with CHARA using MIRC-X/MYSTIC and the future role of nulling.

\subsection{Physics Frontiers Interferometer (new PFI)}
In \S\ref{sec:ras}, we outlined the appetite for a few-element, large-aperture, near-infrared/visible interferometer to answer fundamental questions in a range of astrophysics areas.  We might recast PFI as the Physics Frontiers Interferometer, a different kind of facility that would more affordable than the 12x8m PFI facility and with equally compelling (though different) science goals. 

\section{Summary}
We outlined the science case for the Planet Formation Imager (PFI), and how a 12x8m mid-infrared interferometer would revolutionize planet formation studies.  We also considered a few new project ideas. The outstanding success of the VLTI-GRAVITY shows the power of sensitivity using the 4x8m VLTI-UT array as a PFI pathfinder.  New VLTI projects could address some of the initial PFI science goals (such as finding young giant exoplanets) while also pursuing a range of new topics.  Lastly, we outlined some new activities in space interferometry.

\acknowledgments % equivalent to \section*{ACKNOWLEDGMENTS}       
We would like to acknowledge the nearly 100 scientists and engineers that have contributed to the PFI Project (see [\citenum{pfi2018}] and [\citenum{monnier2018_pfi}] for more complete author lists). This report only reviews recent developments.  

% References
\bibliography{report} % bibliography data in report.bib

\begin{thebibliography}{10}

\bibitem{pfimonnier2014}
{Monnier}, J.~D., {Kraus}, S., {Buscher}, D., {Berger}, J.-P., {Haniff}, C., {Ireland}, M., {Labadie}, L., {Lacour}, S., {Le Coroller}, H., {Petrov}, R.~G., {Pott}, J.-U., {Ridgway}, S., {Surdej}, J., {ten Brummelaar}, T., {Tuthill}, P., and {van Belle}, G., ``{Planet formation imager (PFI): introduction and technical considerations},'' in [{\em Optical and Infrared Interferometry IV}{\nolinebreak\hspace{0.1em}]},  {\em \procspie} {\bf 9146},  914610 (July 2014).

\bibitem{pfikraus2014}
{Kraus}, S., {Monnier}, J., {Harries}, T., {Dong}, R., {Bate}, M., {Whitney}, B., {Zhu}, Z., {Buscher}, D., {Berger}, J.-P., {Haniff}, C., {Ireland}, M., {Labadie}, L., {Lacour}, S., {Petrov}, R., {Ridgway}, S., {Surdej}, J., {ten Brummelaar}, T., {Tuthill}, P., and {van Belle}, G., ``{The science case for the Planet Formation Imager (PFI)},'' in [{\em Optical and Infrared Interferometry IV}{\nolinebreak\hspace{0.1em}]},  {\em \procspie} {\bf 9146},  914611 (July 2014).

\bibitem{pfiireland2014}
{Ireland}, M.~J. and {Monnier}, J.~D., ``{A dispersed heterodyne design for the planet formation imager},'' in [{\em Optical and Infrared Interferometry IV}{\nolinebreak\hspace{0.1em}]},  {\em \procspie} {\bf 9146},  914612 (July 2014).

\bibitem{pfimonnier2016}
{Monnier}, J.~D., {Ireland}, M.~J., {Kraus}, S., {Baron}, F., {Creech-Eakman}, M., {Dong}, R., {Isella}, A., {Merand}, A., {Michael}, E., {Minardi}, S., {Mozurkewich}, D., {Petrov}, R., {Rinehart}, S., {ten Brummelaar}, T., {Vasisht}, G., {Wishnow}, E., {Young}, J., and {Zhu}, Z., ``{Architecture design study and technology road map for the Planet Formation Imager (PFI)},'' in [{\em Optical and Infrared Interferometry and Imaging V}{\nolinebreak\hspace{0.1em}]},  {\em \procspie} {\bf 9907},  99071O (Aug. 2016).

\bibitem{pfikraus2016}
{Kraus}, S., {Monnier}, J.~D., {Ireland}, M.~J., {Duch{\^e}ne}, G., {Espaillat}, C., {H{\"o}nig}, S., {Juhasz}, A., {Mordasini}, C., {Olofsson}, J., {Paladini}, C., {Stassun}, K., {Turner}, N., {Vasisht}, G., {Harries}, T.~J., {Bate}, M.~R., {Gonzalez}, J.-F., {Matter}, A., {Zhu}, Z., {Panic}, O., {Regaly}, Z., {Morbidelli}, A., {Meru}, F., {Wolf}, S., {Ilee}, J., {Berger}, J.-P., {Zhao}, M., {Kral}, Q., {Morlok}, A., {Bonsor}, A., {Ciardi}, D., {Kane}, S.~R., {Kratter}, K., {Laughlin}, G., {Pepper}, J., {Raymond}, S., {Labadie}, L., {Nelson}, R.~P., {Weigelt}, G., {ten Brummelaar}, T., {Pierens}, A., {Oudmaijer}, R., {Kley}, W., {Pope}, B., {Jensen}, E.~L.~N., {Bayo}, A., {Smith}, M., {Boyajian}, T., {Quiroga-Nu{\~n}ez}, L.~H., {Millan-Gabet}, R., {Chiavassa}, A., {Gallenne}, A., {Reynolds}, M., {de Wit}, W.-J., {Wittkowski}, M., {Millour}, F., {Gandhi}, P., {Ramos Almeida}, C., {Alonso Herrero}, A., {Packham}, C., {Kishimoto}, M., {Tristram}, K.~R.~W., {Pott}, J.-U., {Surdej}, J., {Buscher}, D., {Haniff},
  C., {Lacour}, S., {Petrov}, R., {Ridgway}, S., {Tuthill}, P., {van Belle}, G., {Armitage}, P., {Baruteau}, C., {Benisty}, M., {Bitsch}, B., {Paardekooper}, S.-J., {Pinte}, C., {Masset}, F., and {Rosotti}, G., ``{Planet Formation Imager (PFI): science vision and key requirements},'' in [{\em Optical and Infrared Interferometry and Imaging V}{\nolinebreak\hspace{0.1em}]},  {\em \procspie} {\bf 9907},  99071K (Aug. 2016).

\bibitem{pfiireland2016}
{Ireland}, M.~J., {Monnier}, J.~D., {Kraus}, S., {Isella}, A., {Minardi}, S., {Petrov}, R., {ten Brummelaar}, T., {Young}, J., {Vasisht}, G., {Mozurkewich}, D., {Rinehart}, S., {Michael}, E.~A., {van Belle}, G., and {Woillez}, J., ``{Status of the Planet Formation Imager (PFI) concept},'' in [{\em Optical and Infrared Interferometry and Imaging V}{\nolinebreak\hspace{0.1em}]},  {\em \procspie} {\bf 9907},  99071L (Aug. 2016).

\bibitem{pfiminardi2016}
{Minardi}, S., {Lacour}, S., {Berger}, J.-P., {Labadie}, L., {Thomson}, R.~R., {Haniff}, C., and {Ireland}, M., ``{Beam combination schemes and technologies for the Planet Formation Imager},'' in [{\em Optical and Infrared Interferometry and Imaging V}{\nolinebreak\hspace{0.1em}]},  {\em \procspie} {\bf 9907},  99071N (Aug. 2016).

\bibitem{pfimozurkewich2016real}
{Mozurkewich}, D., {Young}, J., and {Ireland}, M., ``{Practical beam transport for PFI},'' in [{\em Optical and Infrared Interferometry and Imaging V}{\nolinebreak\hspace{0.1em}]},  {\em \procspie} {\bf 9907},  99073X (Aug. 2016).

\bibitem{pfibesser2016}
{Besser}, F.~E., {Rates}, A., {Ortega}, N., {Pina}, M.~I., {Pollarolo}, C., {Jofre}, M., {Ya{\~n}ez}, C., {Lasen}, M., {Ramos}, N., and {Michael}, E.~A., ``{Fiber-based heterodyne infrared interferometry: an instrumentation study platform on the way to the proposed Infrared Planet Formation Imager},'' in [{\em Optical and Infrared Interferometry and Imaging V}{\nolinebreak\hspace{0.1em}]},  {\em \procspie} {\bf 9907},  99072L (Aug. 2016).

\bibitem{pfipetrov2016}
{Petrov}, R.~G., {Boskri}, A., {Elhalkouj}, T., {Monnier}, J., {Ireland}, M., and {Kraus}, S., ``{Co-phasing the planet formation imager},'' in [{\em Optical and Infrared Interferometry and Imaging V}{\nolinebreak\hspace{0.1em}]},  {\em \procspie} {\bf 9907},  99073W (Aug. 2016).

\bibitem{pfi2018}
{Monnier}, J.~D., {Ireland}, M., {Kraus}, S., {Alonso-Herrero}, A., {Bonsor}, A., {Baron}, F., {Bayo}, A., {Berger}, J.-P., {Boyajian}, T., {Chiavassa}, A., {Ciardi}, D., {Creech-Eakman}, M., {de Wit}, W.-J., {Defr{\`e}re}, D., {Dong}, R., {Duch{\^e}ne}, G., {Espaillat}, C., {Gallenne}, A., {Gandhi}, P., {Gonzalez}, J.-F., {Haniff}, C., {Hoenig}, S., {Ilee}, J., {Isella}, A., {Jensen}, E., {Juhasz}, A., {Kane}, S., {Kishimoto}, M., {Kley}, W., {Kral}, Q., {Kratter}, K., {Labadie}, L., {Lacour}, S., {Laughlin}, G., {Le Bouquin}, J.-B., {Michael}, E., {Meru}, F., {Millan-Gabet}, R., {Millour}, F., {Minardi}, S., {Morbidelli}, A., {Mordasini}, C., {Morlok}, A., {Mozurkewich}, D., {Nelson}, R., {Olofsson}, J., {Oudmaijer}, R., {Packham}, C., {Paladini}, C., {Panic}, O., {Petrov}, R., {Pope}, B., {Pott}, J.-U., {Quiroga-Nunez}, L.~H., {Ramos Almeida}, C., {Raymond}, S.~N., {Regaly}, Z., {Reynolds}, M., {Ridgway}, S., {Rinehart}, S., {Schreiber}, M., {Smith}, M., {Stassun}, K., {Surdej}, J., {ten Brummelaar}, T.,
  {Tristram}, K., {Turner}, N., {Tuthill}, P., {van Belle}, G., {Vasisht}, G., {Wallace}, A., {Weigelt}, G., {Wishnow}, E., {Wittkowski}, M., {Wolf}, S., {Young}, J., {Zhao}, M., {Zhu}, Z., and {Z{\'u}{\~n}iga-Fern{\'a}ndez}, S., ``{Planet formation imager: project update},'' in [{\em Optical and Infrared Interferometry and Imaging VI}{\nolinebreak\hspace{0.1em}]},  {Creech-Eakman}, M.~J., {Tuthill}, P.~G., and {M{\'e}rand}, A., eds., {\em Society of Photo-Optical Instrumentation Engineers (SPIE) Conference Series} {\bf 10701},  1070118 (July 2018).

\bibitem{2018pfi1}
{Z{\'u}{\~n}iga-Fern{\'a}ndez}, S., {Bayo}, A., {Olofsson}, J., {Pedrero}, L., {Lobos}, C., {Rozas}, E., {Soto}, N., {Schreiber}, M., {Esc{\'a}rate}, P., {Romero}, C., {Hakobyan}, H., {Cuadra}, J., {Rozas}, C., {Monnier}, J.~D., {Kraus}, S., {Ireland}, M.~J., and {Mardones}, P., ``{NPF: mirror development in Chile},'' in [{\em Ground-based and Airborne Telescopes VII}{\nolinebreak\hspace{0.1em}]},  {Marshall}, H.~K. and {Spyromilio}, J., eds., {\em Society of Photo-Optical Instrumentation Engineers (SPIE) Conference Series} {\bf 10700},  107003X (July 2018).

\bibitem{2018pfi2}
{Michael}, E.~A. and {Besser}, F.~E., ``{Fiber-based infrared heterodyne technology for the PFI: on the possibility of breaking the noise temperature quantum limit with cross-correlation},'' in [{\em Optical and Infrared Interferometry and Imaging VI}{\nolinebreak\hspace{0.1em}]},  {Creech-Eakman}, M.~J., {Tuthill}, P.~G., and {M{\'e}rand}, A., eds., {\em Society of Photo-Optical Instrumentation Engineers (SPIE) Conference Series} {\bf 10701},  107011X (July 2018).

\bibitem{2018pfi3}
{Besser}, F.~E., {Ramos}, N., {Rates}, A., {Ortega}, N., {Sepulveda}, S., {Parvex}, T., {Pi{\~n}a}, M., {Pollarolo}, C., {Jara}, R.~E., {Espinoza}, K.~R., {Rodriguez}, A., {Becerra}, A., {Martin}, P., {Cadiz}, M.~D., {Henriquez}, N.~N., {Carrasco}, J., {San Martin}, C., {Rearte}, C., {Diaz}, J., {Moreno}, E., and {Michael}, E.~A., ``{Fiber-based infrared heterodyne technology for the PFI: development of a prototype test system},'' in [{\em Optical and Infrared Interferometry and Imaging VI}{\nolinebreak\hspace{0.1em}]},  {Creech-Eakman}, M.~J., {Tuthill}, P.~G., and {M{\'e}rand}, A., eds., {\em Society of Photo-Optical Instrumentation Engineers (SPIE) Conference Series} {\bf 10701},  107012L (July 2018).

\bibitem{2018pfi4}
{Pedretti}, E., {Diener}, R., {Shankar Nayak}, A., {Tepper}, J., {Labadie}, L., {Pertsch}, T., {Nolte}, S., and {Minardi}, S., ``{Beam combination schemes and technologies for the Planet Formation Imager},'' in [{\em Optical and Infrared Interferometry and Imaging VI}{\nolinebreak\hspace{0.1em}]},  {Creech-Eakman}, M.~J., {Tuthill}, P.~G., and {M{\'e}rand}, A., eds., {\em Society of Photo-Optical Instrumentation Engineers (SPIE) Conference Series} {\bf 10701},  107012O (July 2018).

\bibitem{monnier2018_pfi}
{Monnier}, J.~D., {Kraus}, S., {Ireland}, M.~J., {Baron}, F., {Bayo}, A., {Berger}, J.-P., {Creech-Eakman}, M., {Dong}, R., {Duch{\^e}ne}, G., {Espaillat}, C., {Haniff}, C., {H{\"o}nig}, S., {Isella}, A., {Juhasz}, A., {Labadie}, L., {Lacour}, S., {Leifer}, S., {Merand}, A., {Michael}, E., {Minardi}, S., {Mordasini}, C., {Mozurkewich}, D., {Olofsson}, J., {Paladini}, C., {Petrov}, R., {Pott}, J.-U., {Ridgway}, S., {Rinehart}, S., {Stassun}, K., {Surdej}, J., {Brummelaar}, T.~t., {Turner}, N., {Tuthill}, P., {Vahala}, K., {van Belle}, G., {Vasisht}, G., {Wishnow}, E., {Young}, J., and {Zhu}, Z., ``{The planet formation imager},'' {\em Experimental Astronomy}~{\bf 46},  517--529 (Dec. 2018).

\bibitem{monnier2019_astro2020}
{Monnier}, J.~D. and {endorsers}, ., ``{Setting the Stage for the Planet Formation Imager},'' {\em arXiv e-prints} ,  arXiv:1907.10663 (July 2019).

\bibitem{lacour2021}
{Lacour}, S., {Wang}, J.~J., {Rodet}, L., {Nowak}, M., {Shangguan}, J., {Beust}, H., {Lagrange}, A.~M., {Abuter}, R., {Amorim}, A., {Asensio-Torres}, R., {Benisty}, M., {Berger}, J.~P., {Blunt}, S., {Boccaletti}, A., {Bohn}, A., {Bolzer}, M.~L., {Bonnefoy}, M., {Bonnet}, H., {Bourdarot}, G., {Brandner}, W., {Cantalloube}, F., {Caselli}, P., {Charnay}, B., {Chauvin}, G., {Choquet}, E., {Christiaens}, V., {Cl{\'e}net}, Y., {Coud{\'e} Du Foresto}, V., {Cridland}, A., {Dembet}, R., {Dexter}, J., {de Zeeuw}, P.~T., {Drescher}, A., {Duvert}, G., {Eckart}, A., {Eisenhauer}, F., {Gao}, F., {Garcia}, P., {Garcia Lopez}, R., {Gendron}, E., {Genzel}, R., {Gillessen}, S., {Girard}, J.~H., {Haubois}, X., {Hei{\ss}el}, G., {Henning}, T., {Hinkley}, S., {Hippler}, S., {Horrobin}, M., {Houll{\'e}}, M., {Hubert}, Z., {Jocou}, L., {Kammerer}, J., {Keppler}, M., {Kervella}, P., {Kreidberg}, L., {Lapeyr{\`e}re}, V., {Le Bouquin}, J.~B., {L{\'e}na}, P., {Lutz}, D., {Maire}, A.~L., {M{\'e}rand}, A., {Molli{\`e}re}, P., {Monnier},
  J.~D., {Mouillet}, D., {Nasedkin}, E., {Ott}, T., {Otten}, G.~P.~P.~L., {Paladini}, C., {Paumard}, T., {Perraut}, K., {Perrin}, G., {Pfuhl}, O., {Rickman}, E., {Pueyo}, L., {Rameau}, J., {Rousset}, G., {Rustamkulov}, Z., {Samland}, M., {Shimizu}, T., {Sing}, D., {Stadler}, J., {Stolker}, T., {Straub}, O., {Straubmeier}, C., {Sturm}, E., {Tacconi}, L.~J., {van Dishoeck}, E.~F., {Vigan}, A., {Vincent}, F., {von Fellenberg}, S.~D., {Ward-Duong}, K., {Widmann}, F., {Wieprecht}, E., {Wiezorrek}, E., {Woillez}, J., {Yazici}, S., {Young}, A., and {Gravity Collaboration}, ``{The mass of {\ensuremath{\beta}} Pictoris c from {\ensuremath{\beta}} Pictoris b orbital motion},'' {\em \aap}~{\bf 654},  L2 (Oct. 2021).

\bibitem{wang2020}
{Wang}, J.~J., {Ginzburg}, S., {Ren}, B., {Wallack}, N., {Gao}, P., {Mawet}, D., {Bond}, C.~Z., {Cetre}, S., {Wizinowich}, P., {De Rosa}, R.~J., {Ruane}, G., {Liu}, M.~C., {Absil}, O., {Alvarez}, C., {Baranec}, C., {Choquet}, {\'E}., {Chun}, M., {Defr{\`e}re}, D., {Delorme}, J.-R., {Duch{\^e}ne}, G., {Forsberg}, P., {Ghez}, A., {Guyon}, O., {Hall}, D. N.~B., {Huby}, E., {Jolivet}, A., {Jensen-Clem}, R., {Jovanovic}, N., {Karlsson}, M., {Lilley}, S., {Matthews}, K., {M{\'e}nard}, F., {Meshkat}, T., {Millar-Blanchaer}, M., {Ngo}, H., {Orban de Xivry}, G., {Pinte}, C., {Ragland}, S., {Serabyn}, E., {Catal{\'a}n}, E.~V., {Wang}, J., {Wetherell}, E., {Williams}, J.~P., {Ygouf}, M., and {Zuckerman}, B., ``{Keck/NIRC2 L'-band Imaging of Jovian-mass Accreting Protoplanets around PDS 70},'' {\em \aj}~{\bf 159},  263 (June 2020).

\bibitem{benisty2021}
{Benisty}, M., {Bae}, J., {Facchini}, S., {Keppler}, M., {Teague}, R., {Isella}, A., {Kurtovic}, N.~T., {P{\'e}rez}, L.~M., {Sierra}, A., {Andrews}, S.~M., {Carpenter}, J., {Czekala}, I., {Dominik}, C., {Henning}, T., {Menard}, F., {Pinilla}, P., and {Zurlo}, A., ``{A Circumplanetary Disk around PDS70c},'' {\em \apjl}~{\bf 916},  L2 (July 2021).

\bibitem{christiaens2024}
{Christiaens}, V., {Samland}, M., {Henning}, T., {Portilla-Revelo}, B., {Perotti}, G., {Matthews}, E., {Absil}, O., {Decin}, L., {Kamp}, I., {Boccaletti}, A., {Tabone}, B., {Marleau}, G.~D., {van Dishoeck}, E.~F., {G{\"u}del}, M., {Lagage}, P.~O., {Barrado}, D., {Caratti o Garatti}, A., {Glauser}, A.~M., {Olofsson}, G., {Ray}, T.~P., {Scheithauer}, S., {Vandenbussche}, B., {Waters}, L.~B.~F.~M., {Arabhavi}, A.~M., {Grant}, S.~L., {Jang}, H., {Kanwar}, J., {Schreiber}, J., {Schwarz}, K., {Temmink}, M., and {{\"O}stlin}, G., ``{MINDS: JWST/NIRCam imaging of the protoplanetary disk PDS 70. A spiral accretion stream and a potential third protoplanet},'' {\em \aap}~{\bf 685},  L1 (May 2024).

\bibitem{baraffe1998}
{Baraffe}, I., {Chabrier}, G., {Allard}, F., and {Hauschildt}, P.~H., ``{Evolutionary models for solar metallicity low-mass stars: mass-magnitude relationships and color-magnitude diagrams},'' {\em \aap}~{\bf 337},  403--412 (Sept. 1998).

\bibitem{spiegel2012}
{Spiegel}, D.~S. and {Burrows}, A., ``{Spectral and Photometric Diagnostics of Giant Planet Formation Scenarios},'' {\em \apj}~{\bf 745},  174 (Feb. 2012).

\bibitem{zhu2015}
{Zhu}, Z., ``{Accreting Circumplanetary Disks: Observational Signatures},'' {\em \apj}~{\bf 799},  16 (Jan. 2015).

\bibitem{dong2015}
{Dong}, R., {Zhu}, Z., and {Whitney}, B., ``{Observational Signatures of Planets in Protoplanetary Disks I. Gaps Opened by Single and Multiple Young Planets in Disks},'' {\em \apj}~{\bf 809},  93 (Aug. 2015).

\bibitem{raymond2017}
{Raymond}, S.~N. and {Izidoro}, A., ``{Origin of water in the inner Solar System: Planetesimals scattered inward during Jupiter and Saturn's rapid gas accretion},'' {\em Icarus}~{\bf 297},  134--148 (Nov. 2017).

\bibitem{walsh2011}
{Walsh}, K.~J., {Morbidelli}, A., {Raymond}, S.~N., {O'Brien}, D.~P., and {Mandell}, A.~M., ``{A low mass for Mars from Jupiter's early gas-driven migration},'' {\em \nat}~{\bf 475},  206--209 (July 2011).

\bibitem{ayliffe2009}
{Ayliffe}, B.~A. and {Bate}, M.~R., ``{Gas accretion on to planetary cores: three-dimensional self-gravitating radiation hydrodynamical calculations},'' {\em \mnras}~{\bf 393},  49--64 (Feb. 2009).

\bibitem{rigliaco2015}
{Rigliaco}, E., {Pascucci}, I., {Duchene}, G., {Edwards}, S., {Ardila}, D.~R., {Grady}, C., {Mendigut{\'{\i}}a}, I., {Montesinos}, B., {Mulders}, G.~D., {Najita}, J.~R., {Carpenter}, J., {Furlan}, E., {Gorti}, U., {Meijerink}, R., and {Meyer}, M.~R., ``{Probing Stellar Accretion with Mid-infrared Hydrogen Lines},'' {\em \apj}~{\bf 801},  31 (Mar. 2015).

\bibitem{teague2018}
{Teague}, R., {Bae}, J., {Bergin}, E.~A., {Birnstiel}, T., and {Foreman-Mackey}, D., ``{A Kinematical Detection of Two Embedded Jupiter-mass Planets in HD 163296},'' {\em \apjl}~{\bf 860},  L12 (June 2018).

\bibitem{pinte2018}
{Pinte}, C., {Price}, D.~J., {M{\'e}nard}, F., {Duch{\^e}ne}, G., {Dent}, W.~R.~F., {Hill}, T., {de Gregorio-Monsalvo}, I., {Hales}, A., and {Mentiplay}, D., ``{Kinematic Evidence for an Embedded Protoplanet in a Circumstellar Disk},'' {\em \apjl}~{\bf 860},  L13 (June 2018).

\bibitem{ruge2014}
{Ruge}, J.~P., {Wolf}, S., {Uribe}, A.~L., and {Klahr}, H.~H., ``{Planet-induced disk structures: A comparison between (sub)mm and infrared radiation},'' {\em \aap}~{\bf 572},  L2 (Dec. 2014).

\bibitem{raymond2006b}
{Raymond}, S.~N., {Mandell}, A.~M., and {Sigurdsson}, S., ``{Exotic Earths: Forming Habitable Worlds with Giant Planet Migration},'' {\em Science}~{\bf 313},  1413--1416 (Sept. 2006).

\bibitem{baruteau2014}
{Baruteau}, C., {Crida}, A., {Paardekooper}, S.-J., {Masset}, F., {Guilet}, J., {Bitsch}, B., {Nelson}, R., {Kley}, W., and {Papaloizou}, J., ``{Planet-Disk Interactions and Early Evolution of Planetary Systems},'' {\em Protostars and Planets VI} ,  667--689 (2014).

\bibitem{davies2014}
{Davies}, M.~B., {Adams}, F.~C., {Armitage}, P., {Chambers}, J., {Ford}, E., {Morbidelli}, A., {Raymond}, S.~N., and {Veras}, D., ``{The Long-Term Dynamical Evolution of Planetary Systems},'' {\em Protostars and Planets VI} ,  787--808 (2014).

\bibitem{kingsley2018}
{Kingsley}, J.~S., {Angel}, R., {Davison}, W., {Neff}, D., {Teran}, J., {Assenmacher}, B., {Peyton}, K., {Martin}, H.~M., {Oh}, C., {Kim}, D., {Pearce}, E., {Rascon}, M., {Connors}, T., {Alfred}, D., {Jannuzi}, B.~T., {Zaritsky}, D., {Christensen}, E., {Males}, J., {Hinz}, P., {Seaman}, R., {Gonzales}, K., and {Adriaanse}, D., ``{An inexpensive turnkey 6.5m observatory with customizing options},'' in [{\em Ground-based and Airborne Telescopes VII}{\nolinebreak\hspace{0.1em}]},  {Marshall}, H.~K. and {Spyromilio}, J., eds., {\em Society of Photo-Optical Instrumentation Engineers (SPIE) Conference Series} {\bf 10700},  107004H (July 2018).

\bibitem{monnier2019_roadmap}
{Monnier}, J., {Aarnio}, A., {Absil}, O., {Anugu}, N., {Baines}, E., {Bayo}, A., {Berger}, J.-P., {Cleeves}, L.~I., {Dale}, D., {Danchi}, W., {de Wit}, W.~J., {Defr{\`e}re}, D., {Domagal-Goldman}, S., {Elvis}, M., {Froebrich}, D., {Gai}, M., {Gandhi}, P., {Garcia}, P., {Gardner}, T., {Gies}, D., {Gonzalez}, J.-F., {Gunter}, B., {Hoenig}, S., {Ireland}, M., {Jorgensen}, A.~M., {Kishimoto}, M., {Klarmann}, L., {Kloppenborg}, B., {Kluska}, J., {Knight}, J.~S., {Kral}, Q., {Kraus}, S., {Labadie}, L., {Lawson}, P., {Le Bouquin}, J.-B., {Leisawitz}, D., {Lightsey}, E.~G., {Linz}, H., {Lipscy}, S., {MacGregor}, M., {Matsuo}, H., {Mennesson}, B., {Meyer}, M., {Michael}, E.~A., {Millour}, F., {Mozurkewich}, D., {Norris}, R., {Ollivier}, M., {Packham}, C., {Petrov}, R., {Pueyo}, L., {Pope}, B., {Quanz}, S., {Ragland}, S., {Rau}, G., {Regaly}, Z., {Riva}, A., {Roettenbacher}, R., {Savini}, G., {Setterholm}, B., {Sewilo}, M., {Smith}, M., {Spencer}, L., {ten Brummelaar}, T., {Turner}, N., {van Belle}, G., {Weigelt}, G.,
  and {Wittkowski}, M., ``{A Realistic Roadmap to Formation Flying Space Interferometry},'' in [{\em Bulletin of the American Astronomical Society}{\nolinebreak\hspace{0.1em}]},   {\bf 51},  153 (Sept. 2019).

\bibitem{ireland2020}
{Ireland}, M.~J., ``{Long-baseline space interferometry for astrophysics: a forward look at scientific potential and remaining technical challenges},'' in [{\em Society of Photo-Optical Instrumentation Engineers (SPIE) Conference Series}{\nolinebreak\hspace{0.1em}]},  {\em Society of Photo-Optical Instrumentation Engineers (SPIE) Conference Series} {\bf 11446},  114461A (Dec. 2020).

\bibitem{lagadec2020}
{Lagadec}, T., {Ireland}, M., {Hansen}, J., {Mathew}, J., {Travouillon}, T., and {Madden}, S., ``{Compact unambiguous differential path-length metrology with dispersed Fabry-Perot laser diodes for a space interferometer array},'' in [{\em Society of Photo-Optical Instrumentation Engineers (SPIE) Conference Series}{\nolinebreak\hspace{0.1em}]},  {\em Society of Photo-Optical Instrumentation Engineers (SPIE) Conference Series} {\bf 11446},  114462F (Dec. 2020).

\bibitem{martinod2021}
{Martinod}, M.-A., {Tuthill}, P., {Gross}, S., {Norris}, B., {Sweeney}, D., and {Withford}, M.~J., ``{Achromatic photonic tricouplers for application in nulling interferometry},'' {\em \ao}~{\bf 60},  D100 (July 2021).

\bibitem{hansen2020}
{Hansen}, J.~T. and {Ireland}, M.~J., ``{A linear formation-flying astronomical interferometer in low Earth orbit},'' {\em \pasa}~{\bf 37},  e019 (May 2020).

\bibitem{dandumont2021}
{Dandumont}, C., {Defr{\`e}re}, D., and {Loicq}, J., ``{Feasibility study of an interferometric Small-Sat to study exoplanets},'' in [{\em Society of Photo-Optical Instrumentation Engineers (SPIE) Conference Series}{\nolinebreak\hspace{0.1em}]},  {\em Society of Photo-Optical Instrumentation Engineers (SPIE) Conference Series} {\bf 11852},  118523U (June 2021).

\bibitem{matsuo2022}
{Matsuo}, T., {Ikari}, S., {Kondo}, H., {Ishiwata}, S., {Nakasuka}, S., and {Yamamuro}, T., ``{High spatial resolution spectral imaging method for space interferometers and its application to formation flying small satellites},'' {\em Journal of Astronomical Telescopes, Instruments, and Systems}~{\bf 8},  015001 (Jan. 2022).

\bibitem{quanz2022}
{Quanz}, S.~P., {Ottiger}, M., {Fontanet}, E., {Kammerer}, J., {Menti}, F., {Dannert}, F., {Gheorghe}, A., {Absil}, O., {Airapetian}, V.~S., {Alei}, E., {Allart}, R., {Angerhausen}, D., {Blumenthal}, S., {Buchhave}, L.~A., {Cabrera}, J., {Carri{\'o}n-Gonz{\'a}lez}, {\'O}., {Chauvin}, G., {Danchi}, W.~C., {Dandumont}, C., {Defr{\'e}re}, D., {Dorn}, C., {Ehrenreich}, D., {Ertel}, S., {Fridlund}, M., {Garc{\'\i}a Mu{\~n}oz}, A., {Gasc{\'o}n}, C., {Girard}, J.~H., {Glauser}, A., {Grenfell}, J.~L., {Guidi}, G., {Hagelberg}, J., {Helled}, R., {Ireland}, M.~J., {Janson}, M., {Kopparapu}, R.~K., {Korth}, J., {Kozakis}, T., {Kraus}, S., {L{\'e}ger}, A., {Leedj{\"a}rv}, L., {Lichtenberg}, T., {Lillo-Box}, J., {Linz}, H., {Liseau}, R., {Loicq}, J., {Mahendra}, V., {Malbet}, F., {Mathew}, J., {Mennesson}, B., {Meyer}, M.~R., {Mishra}, L., {Molaverdikhani}, K., {Noack}, L., {Oza}, A.~V., {Pall{\'e}}, E., {Parviainen}, H., {Quirrenbach}, A., {Rauer}, H., {Ribas}, I., {Rice}, M., {Romagnolo}, A., {Rugheimer}, S.,
  {Schwieterman}, E.~W., {Serabyn}, E., {Sharma}, S., {Stassun}, K.~G., {Szul{\'a}gyi}, J., {Wang}, H.~S., {Wunderlich}, F., {Wyatt}, M.~C., and {LIFE Collaboration}, ``{Large Interferometer For Exoplanets (LIFE). I. Improved exoplanet detection yield estimates for a large mid-infrared space-interferometer mission},'' {\em \aap}~{\bf 664},  A21 (Aug. 2022).

\bibitem{wallace2019}
{Wallace}, A.~L. and {Ireland}, M.~J., ``{The likelihood of detecting young giant planets with high-contrast imaging and interferometry},'' {\em \mnras}~{\bf 490},  502--512 (Nov. 2019).

\bibitem{Defrere:2018}
{Defr{\`e}re}, D., {Absil}, O., {Berger}, J.-P., {Boulet}, T., {Danchi}, W.~C., {Ertel}, S., {Gallenne}, A., {H{\'e}nault}, F., {Hinz}, P., {Huby}, E., {Ireland}, M., {Kraus}, S., {Labadie}, L., {Le Bouquin}, J.-B., {Martin}, G., {Matter}, A., {M{\'e}rand}, A., {Mennesson}, B., {Minardi}, S., {Monnier}, J., {Norris}, B., {Orban de Xivry}, G., {Pedretti}, E., {Pott}, J.-U., {Reggiani}, M., {Serabyn}, E., {Surdej}, J., {Tristram}, K.~R.~W., and {Woillez}, J., ``{The path towards high-contrast imaging with the VLTI: the Hi-5 project},'' {\em ArXiv e-prints}  (Jan. 2018).

\end{thebibliography}
\bibliographystyle{spiebib} % makes bibtex use spiebib.bst

\end{document}